\documentclass[12pt]{spieman}  % 12pt font required by SPIE;
\usepackage{amsmath,amsfonts,amssymb,mathtools}
\usepackage{graphicx}
\usepackage{setspace}
\usepackage{tocloft}
\usepackage{latexsym}
\usepackage{graphicx}
\usepackage{bm}
\usepackage[misc,geometry]{ifsym}
\usepackage{hyperref}
\title{Deep conditional generative models for longitudinal single-slice abdominal computed tomography harmonization}

\author[1*]{Xin Yu}
\author[1]{Qi Yang}
\author[2]{Yucheng Tang}
\author[1]{Riqiang Gao}
\author[2]{Shunxing Bao}
\author[3]{Leon Y. Cai}
\author[1]{Ho Hin Lee}
\author[1,2]{Yuankai Huo}
\author[4]{Ann Zenobia Moore}
\author[4]{Luigi Ferrucci}
\author[1,2,3]{Bennett A. Landman}
\affil[1]{Department of Computer Science, Vanderbilt University, Nashville, TN, USA 37212}
\affil[2]{Department of Electrical and Computer Engineering, Vanderbilt University, Nashville, TN, USA, 37212}
\affil[3]{Department of Biomedical Engineering, Vanderbilt University, Nashville, TN, USA, 37212}
\affil[4]{National Institute on Aging, Baltimore, MD, USA}

\cftpagenumbersoff{figure}
\cftpagenumbersoff{table} 
\begin{document} 
\maketitle

\begin{abstract}

Purpose: Two-dimensional single-slice abdominal computed tomography (CT) provides a detailed tissue map with high resolution allowing quantitative characterization of relationships between health conditions and aging. However, longitudinal analysis of body composition changes using these scans is difficult due to positional variation between slices acquired in different years, which leading to different organs/tissues captured. 

Approach: To address this issue, we propose C-SliceGen, which takes an arbitrary axial slice in the abdominal region as a condition and generates a pre-defined vertebral level slice by estimating structural changes in the latent space. 

Results: Our experiments on 2608 volumetric CT data from two in-house datasets and 50 subjects from the 2015 Multi-Atlas Abdomen Labeling Challenge dataset (BTCV) Challenge demonstrate that our model can generate high-quality images that are realistic and similar. We further evaluate our method's capability to harmonize longitudinal positional variation on 1033 subjects from the Baltimore Longitudinal Study of Aging (BLSA) dataset, which contains longitudinal single abdominal slices, and confirmed that our method can harmonize the slice positional variance in terms of visceral fat area. 

Conclusion: This approach provides a promising direction for mapping slices from different vertebral levels to a target slice and reducing positional variance for single-slice longitudinal analysis. The source code is available at: \url{https://github.com/MASILab/C-SliceGen}.
\end{abstract}

% Include a list of up to six keywords after the abstract
\keywords{Abdominal Slice Generation, Longitudinal
Data Harmonization, Body Composition}

{\noindent \footnotesize\textbf{*}Corresponding author,  \linkable{xin.yu@vanderbilt.edu} }

% Include email contact information for corresponding author
% {\noindent \footnotesize\textbf{*}Fourth author name,  \linkable{myemail@university.edu} }

\begin{spacing}{2}   % use double spacing for rest of manuscript

\section{Introduction}
\label{sect:intro}  % \label{} allows reference to this section
 
Body compositional analysis is an important term to determine an individual's health condition which refers to the percentage of fat, muscle, and bone percentages in the human body \cite{kuriyan2018body}. Studying the change of body composition on aging enables better prognosis and early disease detection for various diseases, such as heart disease\cite{de2011body}, sarcopenia \cite{ribeiro2014sarcopenia}, and diabetes \cite{solanki2015body}. Computed tomography body composition (CTBC) is a widely employed technique for assessing body composition \cite{andreoli2016body}. 
\begin{figure*}[h!]
\includegraphics[width=\textwidth]{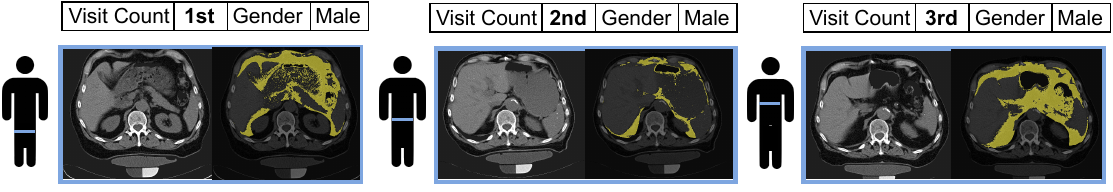}
\caption{An example of a subject with slices acquired at different vertebral levels in different visits. The orange line represents the approximate axial position where the CT scan is taken. The yellow masks represent visceral fat. The shape and size of the captured organs and tissues vary largely among different visits leading to large variations in the visceral fat area.}
\label{fig:fig1}
\end{figure*}
Existing Longitudinal CT scans of the abdomen from the Baltimore Longitudinal Study of Aging (BLSA)~\cite{ferrucci2008baltimore} dataset provide a valuable opportunity to characterize the relationship between
body composition changes and age-related disease, cognitive disease, and metabolic health. To minimize radiation exposure for longitudinal imaging and potential risk associated with contrast administration, 2D non-contrast axial single-slice CT is taken as opposed to 3D volumetric CT commonly acquired in clinical practice. However, it is difficult to locate the same cross-sectional location in longitudinal imaging, and thus there is substantial variation in the organs and tissues captured in different years, as shown in Fig.~\ref{fig:fig1}. The organs and tissues scanned in 2D abdominal slices strongly correlate with body composition measures. Therefore, increased positional variance can make accurately analyzing body composition challenging. Despite this issue, no method has been proposed to address the problem of positional variance in 2D slices. 

 Our goal is to decrease the effects of positional variance in body composition analysis, to facilitate more precise longitudinal interpretation. A major challenge is that the distance between the scans taken in different years is unknown, as the slice can be taken at any abdominal region. Image registration is a commonly used technique in other contexts for correcting pose or positioning errors. However, this approach is not suitable for addressing out-of-plane motion in 2D acquisitions where the tissues/organs that appear in one scan may not appear in the other scan.

Modern generative models based on deep learning have recently shown significant success in generating high-quality and realistic images. The fundamental concept of generative modeling is to train a generative model to learn a distribution so that the generated samples $\hat{x} \sim p_d(\hat{x})$ are from the same distribution as the training data distribution $x \sim p_d(x)$~\cite{bond2021deep}. By learning the joint distribution between the input and target slices, these models can effectively address the limitations of registration. Variational autoencoders (VAEs) \cite{kingma2013auto}, which are a type of generative model, consisting of an encoder and a decoder. The encoder encodes inputs to an interpretable latent distribution, and the decoder decodes the samples of the latent distribution to new data. Generative adversarial networks (GANs)~\cite{NIPS2014_5ca3e9b1} is another type of generative model, which contains two sub-models, a generator model that generates new data and a discriminator that distinguishes between real and generated images. By playing this two-player min-max game, GANs can generate realistic images.   VAEGAN \cite{larsen2016autoencoding} incorporates GAN into VAE framework to create better synthesized images. By using the discriminator to distinguish between real and generated images, VAEGAN can generate more realistic and high-quality images than traditional VAE models. However, original VAEs and GANs suffer from the limitation of lack of control over the generated images. This issue is addressed by conditional GAN (cGAN) \cite{mirza2014conditional} and conditional VAE (cVAE) \cite{sohn2015learning}which allow for generating specific images with a condition, providing more control over the generated outputs. However, the majority of these conditional methods necessitate specific target information such as a target class, semantic map, or heatmap \cite{de2019conditional} as a condition during the testing phase, which is not feasible in our scenario since we do not have any direct target information available.

To provide a condition during testing, we aim to have the network generate a slice of that specific target at a pre-determined vertebral level, which will serve as the generation target. By defining the target slice, the generative model will implicitly learn the organ/tissue composition in the target slice and have this condition learned during training time. We hypothesize that by giving an arbitrary abdominal slice, the model will generate the slice at the target vertebral level while preserving subject-specific information derived from the conditional image such as body habitus. Inspired by \cite{henderson2021unsupervised,de2019conditional,tang2021pancreas}, we introduce the Conditional SliceGen (C-SliceGen) model based on VAEGAN, which enables the generation of subject-specific target vertebral level slices from an arbitrary abdominal slice input. We use 3D volumetric data to train and validate our model since in 3D data the target slice (ground truth) is available for direct comparison with the generated images. The training datasets include an in-house portal venous phase CT and an in-house non-contrast phase CT volume with 1120 and 1488 subjects, respectively. We further evaluate on the 2015 Multi-Atlas Abdomen Labeling Challenge dataset (BTCV) MICCAI Challenge 3D CT dataset \cite{landman2015miccai} for external validation. Structural Similarity
Index (SSIM) \cite{wang2004image}, Peak Signal-to-Noise Ratio (PSNR)\cite{hore2010image}, Learned Perceptual Image Patch Similarity (LPIPS) \cite{zhang2018perceptual}, and Normalized Mutual Information (NML) ~\cite{shannon1948mathematical} are used for image quality assessment. We further apply our trained model on the Baltimore Longitudinal Study of Aging (BLSA) dataset with 1033 subjects to illustrate our model's capability on reducing longitudinal variance caused by positional variation by comparing body composition metrics change before and after the harmonization.

This paper is an extension of our conference version ~\cite{yu2022reducing}. We focus on improving the generalizability of our model. We further apply our model on an in-house non-contrast phase CT dataset to minimize the domain shift problem among different CT phases and revisit the target slice selection method. Second, we add more metrics to evaluate our model and generated images in both 3D datasets and 2D single-slice dataset. Third, we conduct a comprehensive evaluation on the BLSA dataset with 1033 subjects, which is a significant increase compared to the 20 subjects evaluated in the conference version. Last, we conduct an ablation study on validating the most effective distance range between the given and target slice for our proposed method.

Our contributions in this work can be summarized as follow: 
\begin{itemize}
\item[\textbullet] We introduce a VAEGAN-based generative model, C-SliceGen that can generate the subject-specific abdominal slice at a pre-defined target vertebral level by giving a subject arbitrary axial slice in the abdominal region as input. 
\item[\textbullet] Our generative approach can implicitly incorporate the unknown target slice without the information during the testing phase. 
\item[\textbullet] The approach generates realistic and structurally similar images to the true, but unknown, target slices.
\item[\textbullet] Our experiments demonstrate that the proposed method can harmonize variance in body composition metrics caused by positional variation in the longitudinal setting for accurate longitudinal analysis.
\end{itemize}

\section{Method}
\subsection{Technical Background}
\label{problem}
Here, we briefly review the generative models (VAE and GAN) and their variants that our C-SliceGen is based. While generative models have been effective in generating samples, no existing method can be directly applied to our task.
% VAEs is a class of generative model that encode the input to a interpretable latent distribution and allow it to generate new data [].

\subsubsection{VAE}
 VAEs can be expressed probabilistically as $P(x) = P(z)P(x|z)$, where x represents input images and z represents latent variables. The goal of these models is to maximize the likelihood of $p(x)=\int p(z)p_\theta(x|z)dz$, where $z\sim N(0,1)$ is the prior distribution, $p_\theta(x|z)dz$ is the posterior distribution, and $\theta$ represents the decoder parameters. However, it is not feasible to find decoder parameters $\theta$ that maximize the log-likelihood. Instead, VAEs optimize encoder parameters $\phi$ by estimating $p_\theta(x|z)$ using $q_\phi(z|x)$, which is assumed to be a Gaussian distribution with $\mu$ and $\sigma$ as the outputs of the encoder. VAEs is trained by optimizing the Evidence Lower Bound (ELBO).
 
\begin{equation}
L_{VAE}(\theta,\phi,x) = E[\log p_{\theta}(x|z)] - D_{KL}[q_{\phi}(z|x)||p_{\theta}(z)],
\end{equation}
where $E[\log p_{\theta}(x|z)]$ represents the reconstruction loss and the KL-divergence incentivizes the posterior distribution to be close to the prior distribution $p(z)$. During testing, new data can be generated by sampling from the normal distribution $z \sim N(0,1)$ and inputting it into the decoder. Conditional VAE can be optimized with the following ELBO equation as well with little modification:

\begin{equation}
L_{VAE}(\theta,\phi,x,c) = E[\log p_{\theta}(x|z,c)] - D_{KL}[q_{\phi}(z|x,c)||p_{\theta}(z|c)],
\end{equation}

\subsubsection{GAN}

GANs consist of two parts: discriminator and generator. Suppose we have input noise variables $p_z(z)$, the generator will map the input noise to data space $G(z)$ and mix it with the real data $x$. The discriminator $D$, on the other hand, transforms image data into a probability indicating whether the image belongs to the real data distribution or the generator distribution~\cite{creswell2018generative}. To be more specific, the discriminator and the generator play the two-player minmax game with value function $V(D,G)$ in the following manner~\cite{goodfellow2020generative}:
 
% \begin{equation}
% \begin{aligned}
% \min_{G}\max_{D}V(D,G) = &\mathbb{E}_{x\sim p_{\text{data}}(x)}[\log{D(x)}] \\
% &+ \mathbb{E}_{z\sim p_{\text{z}}(z)}[1 - \log{D(G(z))}],
% \end{aligned}
% \end{equation}

\begin{equation}
\min_{G}\max_{D}V(D,G) = \mathbb{E}_{x\sim p_{\text{data}}(x)}[\log{D(x)}] + \mathbb{E}_{z\sim p_{\text{z}}(z)}[1 - \log{D(G(z))}],
\end{equation}

The Wasserstein GAN  with gradient penalty (WGAN-GP)~\cite{gulrajani2017improved} is an alternative to traditional GAN which enhances the stability of the model during training and address problem such as model collapse. The loss function of WGAN-GP can be written as:
\begin{figure*}[h!]
\includegraphics[width=\textwidth]{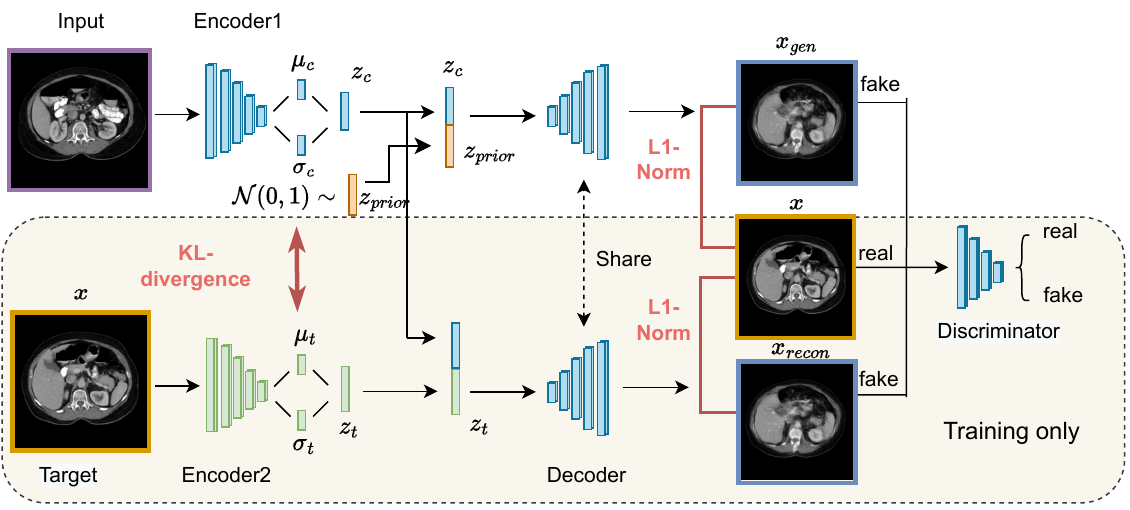}
\caption{ The input image is an arbitrarily acquired slice in the abdominal region. During the training phase, target images (x) are used as the ground truth for the generation and reconstruction process. Latent variables such as $z_c$, $z_t$ and $z_{prior}$ are derived from conditional images, target images, and the normal Gaussian distribution, respectively. $x_{gen}$ and $x_{recon}$ are considered as fake images and target images are considered as real images for the discriminator.}
\label{fig:fig2}
\end{figure*}

% \begin{equation}
% \begin{aligned}
% L_{WGAN-GP} = &\mathbb{E}_{\tilde{x}\sim\mathbb{P}_g}[D(\tilde{x})] +\mathbb{E}_{x\sim \mathbb{P}_r}[D(x)] \\
% &+ \lambda \mathbb{E}_{\hat{x}\sim \mathbb{P}_{\hat{x}}}[(||\bigtriangledown_{\hat{x}}D(\hat{x})||_2 - 1 )^2] ,
% \end{aligned}
% \label{equ:equ2}
% \end{equation}

\begin{equation}
L_{WGAN-GP} = \mathbb{E}_{\tilde{x}\sim\mathbb{P}_g}[D(\tilde{x})] +\mathbb{E}_{x\sim \mathbb{P}_r}[D(x)] 
+ \lambda \mathbb{E}_{\hat{x}\sim \mathbb{P}_{\hat{x}}}[(||\bigtriangledown_{\hat{x}}D(\hat{x})||_2 - 1 )^2] ,
\label{equ:equ2}
\end{equation}
where $\mathbb{P}_g$, $\mathbb{P}_r$ and $\mathbb{P}_{\hat{x}}$ represents the generator distribution, data distribution, and random sample distribution, respectively.
% \subsubsection{cVAEGAN}
% GANs are adversarial generative models contains two sub-models: a generation model that aim to generate new data to fool the discriminator and a discriminator that tried to classify the real and fake data.

\subsection{C-SliceGen}

% In our scenario, the acquired slice can be located at any vertebral level within the abdominal region, and the aim is to generate a new slice at a predetermined target vertebral level using these arbitrary acquired slices. The proposed method is shown in Fig.~\ref{fig:fig2}. The model consists of two encoders, one decoder, and one discriminator. The conditional image for the model is the arbitrary slice, which contains subject-specific information such as organ shape and tissue localization. We assume that this information is preserved and can be interpreted after being encoded to latent variables $z_c$ by encoder1.

Our task involves generating a new slice at a pre-defined vertebral level from an arbitrary slice that is obtained at any vertebral level within the abdominal region. Our proposed method is illustrated in Fig.~\ref{fig:fig2}, which comprises two encoders, one decoder, and one discriminator. 

% In our scenario, acquired slice can be in any vertebral level within the abdominal region. The goal is to use these arbitrary slices to synthesize a new slice at a pre-defined target vertebral level. The overall method is shown in Fig.~\ref{fig:fig2}. There are two encoders, one decoder and one discriminator. The arbitrary slice is the conditional image for the model, which provides subject-specific information such as organs shape and tissue localization. We assume this information remains interpretable after encoding to latent variables $z_c$ by encoder1.
% The target images is the ground truth for the generation, which only exist in the training phase, which . The target and conditional are encoded by two different encoders into the latent variables $z_t$ and $z_c$.
% which  

\subsection{Target Slice Selection}
\label{problem}
The first step is to select a target slice for each individual, with the criterion of selecting a slice that is most similar in terms of organ/tissue structure and appearance across all subjects. Choosing comparable target slices for each individual is a challenging task as it involves taking into account subject-specific variations in organ structure and body composition. We use two methods to select similar target slices for each subject:

\indent {\bf BPR-based Method} We select slices with similar body part regression (BPR) score \cite{tang2021body} as the target slices across subjects. BPR gives different scores to different slices in the abdominal region and is efficient in locating slices. We first select a target slice in a reference subject and document its BPR score, and then we select the slices that have the most similar BPR score as the target slices across subjects.

\indent {\bf Registration-based Method} Initially, we select a reference subject's slice as the reference target slice. Subsequently, we register the axial slices of every subject's volume to the reference target slice and identify the slice that has the largest NML score as the subject target slice.

\subsubsection{Training}
 % \noindent {\bf Training}
The input image for the model is the arbitrary slice, which provides subject-specific information, including organ shape and tissue localization. Note the assumption that the input is intended to represent the target which is not a random input. We believe that this information remains interpretable after encoding to latent variables $z_c$ by encoder1.
All selected target slices ($x$) should have similar organ/tissue structures and appearances. This information is encoded in the latent variables $z_t$. The distribution of $z_t$ can be expressed as $q_\phi(z_t|x)$, where $\phi$ denotes the encoder2 parameters. We combine the organ/tissue structure and appearance of the target slice with the subject-specific information by concatenating the latent variables zc and zt. This combination facilitates the decoder to reconstruct the target slice for the given individual. To regularize the reconstruction process, we compute the L1-norm between the target slice ($x$) and the reconstructed slice ($x_{recon}$) using the following equation:
\begin{equation}
L_{recon} = \lVert\bm{x} - \bm{x}_{recon}\rVert,
\end{equation}
Since no target slice is available during testing, we follow the similar approach as in VAEs. We assume that $q_\phi(z_t|x)$ is a Gaussian distribution with parameters $\mu_t$ and $\sigma_t$, which are the outputs of encoder2. We optimize the KL-divergence to encourage $q_\phi(z_t|x)$ to be close to the prior distribution $z_{prior} \sim N(0,1)$, which can be written as:

\begin{equation}
L_{KL}=\frac{1}{2}\sum_{k=1}^K(1+\log(\sigma_k^2) - \mu_k^2 - \sigma_k^2),
\end{equation}
where $K$ represents the dimension of the latent space. To mimic the process of image generation in the testing phase, we added another input to the decoder by concatenating  $z_c$ with $z_{prior}$ for target slices generation. We denote these generated images as $x_{gen}$. $x_{gen}$ is also regularized by L1-Norm with equation: 
\begin{equation}
L_{gen} = \lVert\bm{x} - \bm{x}_{gen}\rVert,
\end{equation}
The combined loss function of the above-mentioned steps can be expressed as:
\begin{equation}
L_{cVAE}=L_{recon} + L_{gen} + L_{KL},
\label{equ:equ4}
\end{equation}

However, the major drawback of VAEs is that it tends to generate blurry images. On the other hand, GANs can produce images with sharp edges. Following \cite{larsen2016autoencoding}, we add GAN regularization into our model. The discriminator in our proposed C-SliceGen model classifies both the generated and reconstructed images as fake images, while the target images are considered real images. The decoder acts as the generator for the GAN part. GAN loss adds another constraint to force the generated and target images to be similar. The total loss function for our C-SliceGen model can be written as follow:
\begin{equation}
L = L_{cVAE} + \beta L_{GAN},
\label{equ:equ5}
\end{equation}
The adversarial regularization is adjustable by the weighting factor $\beta$.

% However, maximize likelihood function is inherently a difficult problem which can cause blurry generated images. GANs on the other hand increase the image quality in an adversarial manner. Following \cite{larsen2016autoencoding}, we combine GAN with our VAE model. The generated image and reconstructed image both serve as fake images, and the target images serve as the real images for the discriminator to perform classification. The decoder serves as the generator. By sharing the same parameters for the generated and reconstructed images, the GAN loss adds another constraint to force them to be similar. The total loss function of our proposed C-SliceGen can be written as:
% % $L = L_{cVAE} + \beta L_{GAN}$, 
% \begin{equation}
% L = L_{cVAE} + \beta L_{GAN},
% \label{equ:equ5}
% \end{equation}
% where $\beta$ is a weighting factor that determines the adversarial regularization. 

% \noindent {\bf Testing} 
\subsubsection{Testing}
During testing, encoded conditional image $z_c$ is combined with $z_{prior}$, which is sampled from a normal Gaussian distribution, and then feed into the decoder to generate the target slice.

\section{Implementation Details}
\subsection{Dataset}
We train and evaluate our methods on 3D volumetric CT datasets in both the portal venous phase and the non-contrast phase as well as 2D single-slice CT dataset in the non-contrast phase. 
\label{dataset}

\indent\textbf{In-house Portal Venous Dataset} This dataset contains 1120 3D Portal Venous CT volumes from 1120 de-identified subjects from Vanderbilt University Medical Center (VUMC). The data has been approved by the Institutional Review Board (IRB) with IRB \#160764. Quality check is performed on every CT scan to ensure normal abdominal anatomy. The dataset is divided into training, validation, and testing with 1029, 8, and 83 subjects, respectively.

\indent\textbf{In-house Non-contrast Dataset} To minimize the domain shift problem when applying models trained with the portal venous dataset to non-contrast single-slice data, we further train and validate our method using a 3D non-contrast CT dataset with IRB \#172167. This dataset contains 1488 subjects. We split the dataset into training, validation, and testing with 1059, 117, and 312 subjects, respectively.

\indent\textbf{BTCV Dataset} The MICCAI 2015 Multi-Atlas Abdomen Labeling Challenge (BTCV) dataset consisting of 30 Portal Venous CT volumes for training and 20 for testing, is used for the evaluations. We finetune the model trained with the in-house portal venous dataset with 22 data for training and 8 for validation.

\indent\textbf{BLSA Dataset} We assess the effectiveness of our method in harmonizing the positional variation on single-slice data with the Baltimore Longitudinal Study of Aging (BLSA) dataset. To minimize the radiation exposure during longitudinal imaging, the BLSA CT protocol captures single-slice data at specific anatomical landmarks instead of acquiring 3D CT data as is typically done in clinical settings. A total of 1033 subjects have more than 1 visit with some subjects having up to 12 visits and the median number of scans being 3 for the past 15 years. The total number of CT axial scans is 4223, and all the scans are in the non-contrast phase.

% A total of 1120 subjects from Vanderbilt University Medical Center (VUMC) are 
% The models are trained and validated on a large dataset containing 1170 3D Portal Venous CT volumes from 1170 de-identified subjects under Institutional Review Board (IRB) protocols. Each CT scan is quality checked for normal abdominal anatomy. The evaluations are performed on the MICCAI 2015 Multi-Atlas Abdomen Labeling Challenge dataset which contains 30 and 20 abdominal Portal Venous CT volumes for training and testing, respectively. We further evaluate our method's efficacy on reducing positional variance for longitudinal body composition analysis on 20 subjects from the BLSA non-contrast single slice CT dataset. Each subject has either 2 or 3 visits for the past 15 years. 

% \subsection{Implementation Details}
% 
\subsection{Metrics}
We perform a quantitative evaluation of our C-SliceGen generative models using different target slice selection approaches and varying values of $\beta$ (as defined in Equation 5).

\indent {\bf Metrics for 3D Datasets} For the 3D volumetric dataset where the target slices are available for comparison, we use four metrics to assess image quality: SSIM, PSNR, LPIPS, and NML. SSIM assesses the image based on three factors: luminance, contrast, and
structure, and the final score is derived from the multiplication of those three independent factors. PSNR is most determined by mean squared error (MSE). LPIPS is utilized to assess the perceptual similarity of two
images by calculating the similarity between the activations of their respective patches through a pre-defined network. This measure is highly correlated with human perception. NMI measures the degree of information present in one image that contains in the other image~\cite{maes2015image}.

\indent{\bf Metrics for 2D Single-slice Dataset} In the BLSA dataset, each subject only has one axial abdominal CT scan taken per visit, resulting in the absence of ground truth (GT) for direct comparison with the generated target slices. We use NML and coefficient of variation (CV) to evaluate the model performance on harmonizing the positional variation. CV indicates the amount of differences between scans~\cite{wittens2021inter,yu2023longitudinal}, which is defined as:
\begin{equation}
CV = \frac{\sigma}{\mu},
\label{equ:cv}
\end{equation}

% We quantitatively evaluate our generative models C-SliceGen with different $\beta$ (Eq.~\ref{equ:equ5}) and the two different target slice selection approaches using three image quality assessment metrics: Structural Similarity
% Index (SSIM) \cite{wang2004image}, Peak Signal-to-Noise Ratio (PSNR)\cite{hore2010image}, and Learned Perceptual Image Patch Similarity (LPIPS) \cite{zhang2018perceptual}. 

% Although SSIM and PSNR are widely used image quality assessment metrics, the rather simple and shallow function is hard to capture the underlaying perceptual similarity between two images, which leads to a higher score for the blurry images. LPIPS, on the other hand, shows better ability on assessing the perceptual similarity between a high resolution and a superresolved image.

 % \noindent {\bf Training \& Testing}
\subsection{Training \& Testing} 
BPR~\cite{tang2021body} is used to ensure a consistent field of view (FOV) in the abdominal region for all the 3D volumetric data. We preprocessed the data with a soft-tissue CT window range of [-125, 275] Hounsfield Units (HU) and further rescale the data to the range of 0 to 1. The 2D axial CT scans are resized from size 512 $\times$ 512 to 256 $\times$ 256 before being fed into the models. Pytorch is used to implement the proposed methods, with the Adam optimizer and a learning rate of 1e-4, and a weight decay of 1e-4 to optimize the network's total loss when training the model from scratch. When finetuning on the BTCV dataset, the learning rate is reduced to 1e-5. The encoder, decoder, and discriminator structures are modified based on \cite{gao2021lung,li2020learning}. We adopt common data augmentation methods such as shift, rotation, and flip with a probability of 0.5 to facilitate training. 

% The encoder, decoder, and discriminator structures in Gao et al. (2021) and Li and Marlin (2020) are adapted to fit the input size. Online data augmentation is applied using shift, rotation, and flip techniques. The testing phase uses 2614 slices from 83 subjects in-house, and the results are presented in Table 1.

% All 3D volumes undergo BPR to ensure a similar Field of View (FOV). The 2D axial CT scans have image sizes of $512 \times 512$ and are resized to $256 \times{256}$ before feeding into the models.The data are processed with soft-tissue CT window range [-125, 275] HU and rescaled to [0.0,1.0] to facilitate training. The proposed methods are implemented using Pytorch. We use Adam optimizer with a learning rate of $1e-4$ and weight decay of $1e-4$ to optimize the total loss of the network. We adapt the encoder, decoder and discriminator structures in \cite{gao2021lung,li2020learning} to fit our input size. Shift, rotation and flip are used for the online data augmentation.
\begin{table*}[ht]
\centering
\caption{Quantitative results on two in-house test set and BTCV test set using different target slice selection method with different $\beta$ in Eq.~\ref{equ:equ5} for training.}
\label{tab:my-table}
\begin{tabular}{p{4.5cm}|p{2.5cm}p{2.5cm}p{2.5cm}p{2.5cm}}
\hline
Method  & SSIM $\uparrow$ & PSNR  $\uparrow$ & LPIPS $\downarrow$ & NML $\uparrow$ \\ \hline
\multicolumn{5}{c}{the In-house Portal Venous Dataset} \\ \hline
$\beta = 0$, Registration & \textbf{0.636} & \textbf{17.634} & 0.361 & 0.344 \\
$\beta = 0$, BPR & 0.618 & 16.470 & 0.381 & 0.316 \\
$\beta = 0.01$, Registration & 0.615 & 17.256  & \textbf{0.209} & \textbf{0.377}\\
$\beta = 0.01$, BPR & 0.600 & 16.117 & 0.226 & 0.364\\ \hline
 \multicolumn{5}{c}{the BTCV Dataset} \\ \hline
$\beta = 0$, Registration & 0.603 & 17.367 & 0.362 & 0.340\\
$\beta = 0$, BPR & \textbf{0.605} & \textbf{17.546} & 0.376 & 0.330 \\
$\beta = 0.01$, Registration & 0.583 & 16.778 & \textbf{0.208} & 0.422\\
$\beta = 0.01$, BPR & 0.588 & 16.932 & 0.211 & \textbf{0.423}\\ \hline
 \multicolumn{5}{c}{the In-house Non-contrast Dataset} \\  \hline
$\beta = 0.01$, semi-BPR& 0.570 & 18.312 & 0.210 & 0.405 \\ \hline

\end{tabular}
\end{table*} 
% 2614 slices from in-house 83 subjects are used for testing. The results are shown in Table.~\ref{tab:my-table}.
\begin{figure*}[h!]
\centering
\includegraphics[width=\textwidth]{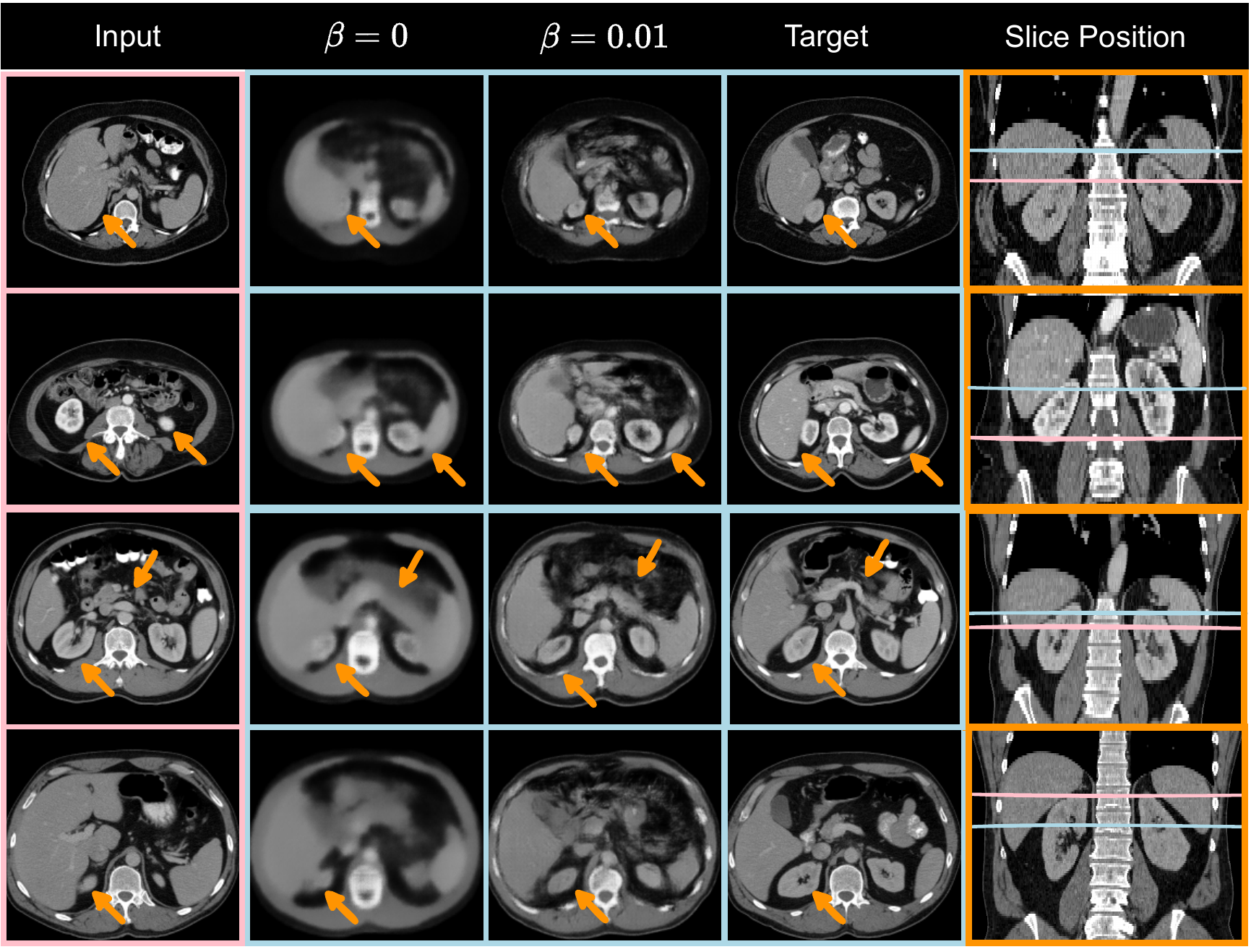}
\caption{The image enclosed within a pink bounding box depicts the input slice, while the image enclosed within a light blue bounding box represents the model outputs and target slice from four different subjects in the BTCV test set. The pink and light blue lines in the rightmost column indicate the axial position of the input and target slices, respectively. The results indicate our model can implicitly learn the subject-specific target slices and generate realistic and structurally similar slices given input slices from arbitrary vertebral levels. }
\label{fig3}
\end{figure*} 
\section{Results}
\subsection{3D Datasets Evaluation}
% \noindent {\bf BTCV Evaluation}
We present the quantitative performance of our model with various metrics on different datasets in Table~\ref{tab:my-table}. Comparing with target slices selected by the registration-based method and BPR-based method, the registration-based method achieve better performance on the in-house portal venous dataset while the BPR-based method performs slightly better on the BTCV dataset which might indicate that the BPR-based method and registration-based method have comparable performance on selecting target slices on the portal venous phase CT scans. We show qualitative results on the BTCV test set in Fig.~\ref{fig3}, which demonstrates that our model is capable of generating target slices irrespective of whether the conditional slice is at a higher, lower, or similar vertebral level.
\begin{figure*}[h!]
\centering
\includegraphics[width=0.75\textwidth]{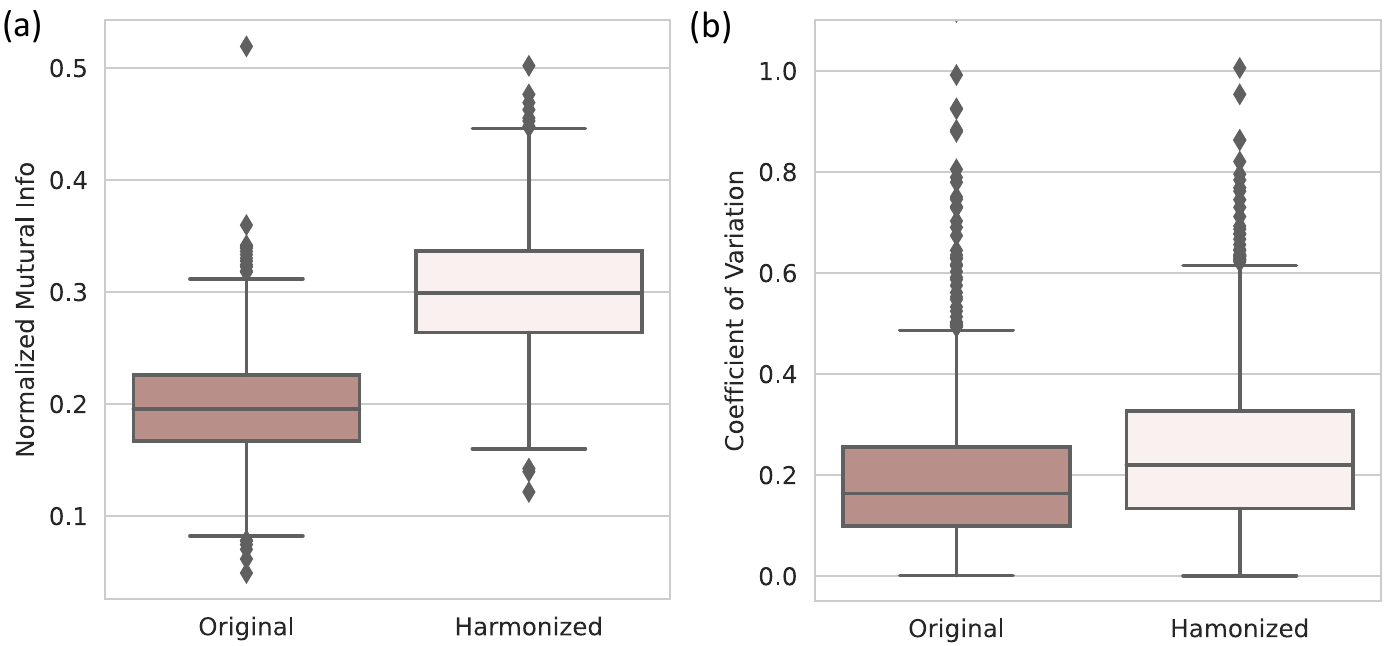}
\caption{Quantitative results of applying the trained model on 1033 subjects from the BLSA dataset. (a) represents the results with NML as metrics, (b) represents the results with CV as metrics. We observed higher NML and CV among the harmonized images. NML and CV show contradictory results where higher NML suggests that the generated images of the same subjects share a higher degree of similar information compared to those of the original images while higher CV implies the differences between slices are increased after harmonization.}
\label{fig:ml}
\end{figure*}
On the non-contrast dataset, however, our empirical results indicate that the generated images are not on a similar vertebral level as opposed to what they suppose to be. We trace back the reason and find that the selected target slices in the non-contrast dataset are not on a similar vertebral level initially, making the network hard to learn the target location and resulting in significant noise in the generated images. To address this issue, we use a semi-BPR method wherein we compare the target slice with the eight axial slices preceding and succeeding it to select the new target slice. The results with the manually corrected target slices are present in Table~\ref{tab:my-table}. Comparing the results from the non-contrast phase dataset and the portal venous phase dataset, two out of four metrics show slightly worse performance while the other two metrics show comparable or even better performance. This may indicate our model has good generalizibility on different CT phases.

% Before evaluating on the BTCV test set, the models are fine-tuned on the BTCV training set to minimize the dataset domain gap with the same training settings except that the learning rate is reduced to $1e-5$. The split of train/validation/test is 22/8/20. The quantitative results and qualitative results are shown in Table.~\ref{tab:my-table} and Fig.~\ref{fig3}, respectively.

% \noindent {\bf BLSA Evaluation}
\subsection{2D Single-slice Evaluation}
We use NML and CV to evaluate our model's longitudinal variation harmonization capability. We evaluate the model's performance using the visceral fat area as the primary metric. This is because visceral fat is highly susceptible to positional variation as mentioned in ~\cite{yu2023longitudinal}, and it is a crucial component of body composition which indicates an individual's health 
condition~\cite{yu2023longitudinal,duren2008body}.
\begin{figure}
\centering
\includegraphics[width=0.35\textwidth]{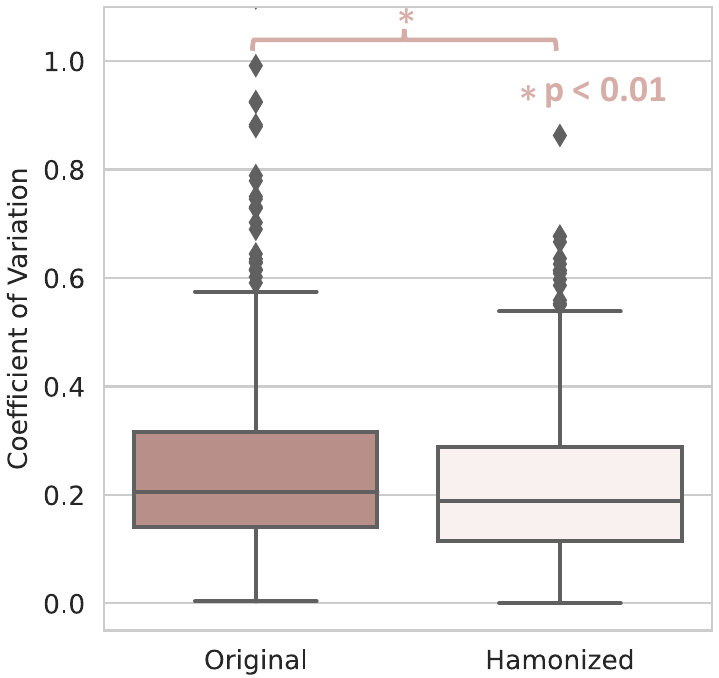}
\caption{CV result of applying the trained model on subjects that have at least one scan that is taken in obvious different vertebral levels. Note: * represents statistically significant ($p < 0.01$) by Wilcoxon signed-rank test. The result demonstrates that our model is effective in harmonizing the variance caused by positional variation in longitudinal imaging.  }
\label{fig:cv}
\end{figure}

The BLSA single-slices CT scans are fed into the model trained with non-contrast 3D CT volumes. The generated images are resized to the original image size of $512 \times 512$. We use the method in ~\cite{yu2023longitudinal} to extract the segmentation mask of the visceral fat which includes feeding the data in pre-trained model for inner/outer abdominal wall segmentation and using fuzzy c-means 
\cite{bezdek1984fcm,tang2020prediction} to extract the adipose tissues. 
In the inner abdominal wall segmentation, we observe that the model performs unsatisfactory to exclude the retroperitoneum from both real and generated images. The retroperitoneum is an anatomical region situated behind the abdominal cavity, which comprises the aorta, and left and right kidneys, and often lacks well-defined boundaries, making it difficult to segment accurately. We follow the practice in ~\cite{yu2023longitudinal} and manually assess the results from both the real and generated images to ensure that the retroperitoneum is segmented correctly. We mask the inner abdominal wall with the adipose tissues to get the final visceral fat mask.

\begin{figure*}[h!]
\centering
\includegraphics[width=\textwidth]{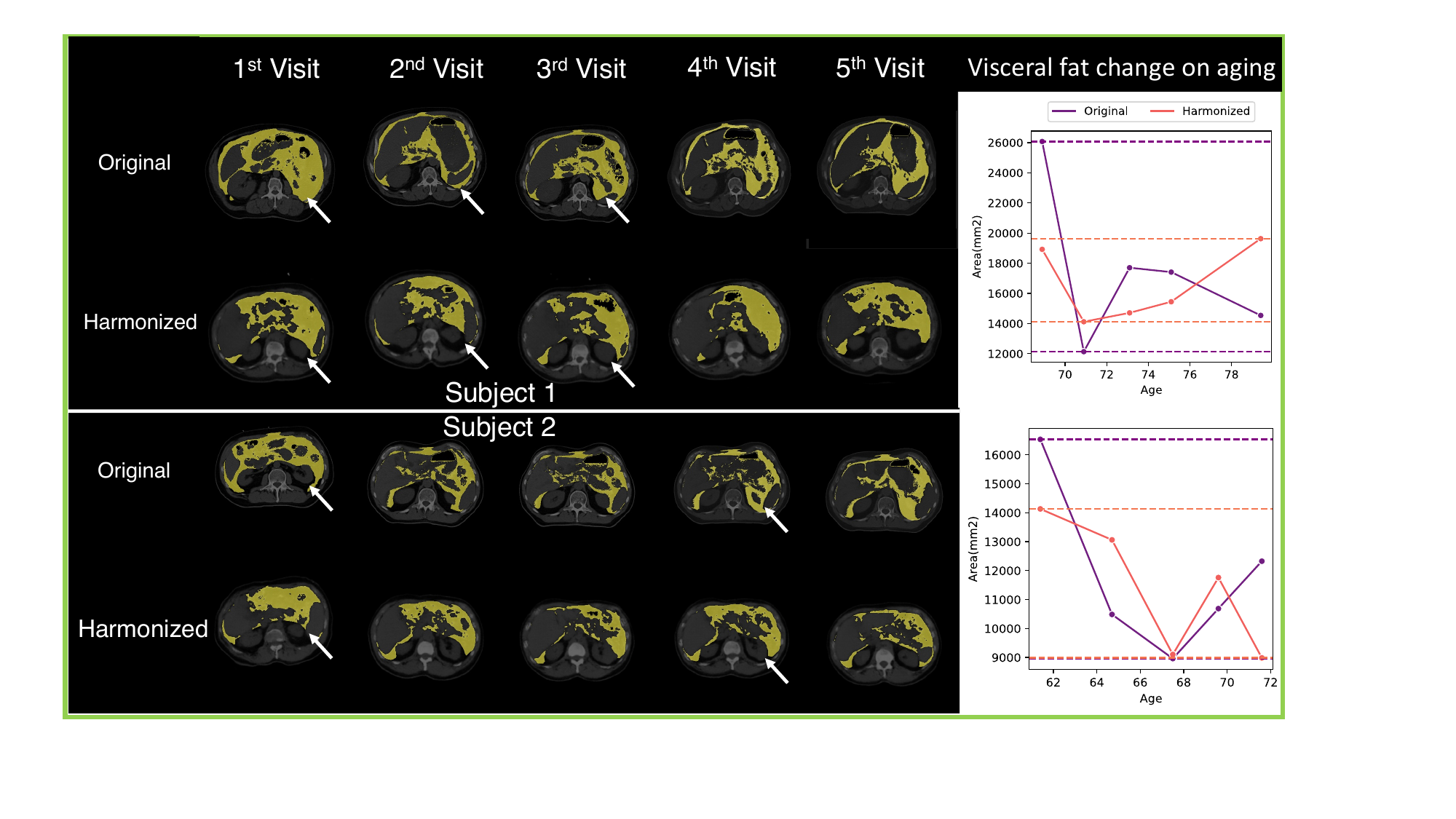}
\caption{The yellow mask represents the visceral fat for two subjects in the BLSA dataset that have 5 repeated visits. The rightmost column shows the visceral fat area change on aging. The horizontal lines represent the maximum and minimum values of different image types. We highlight the images with different organs or tissues captured compared to most other images by the white arrows. The generated images manage to harmonize organ/tissues difference. We observed the range between the maximum and minimum value of the among different visits decreased after harmonization which implies the harmonized images have less variation.}
\label{fig:blsa_vis_short}
\end{figure*}

\begin{figure*}[h!]
\centering
\includegraphics[width=\textwidth]{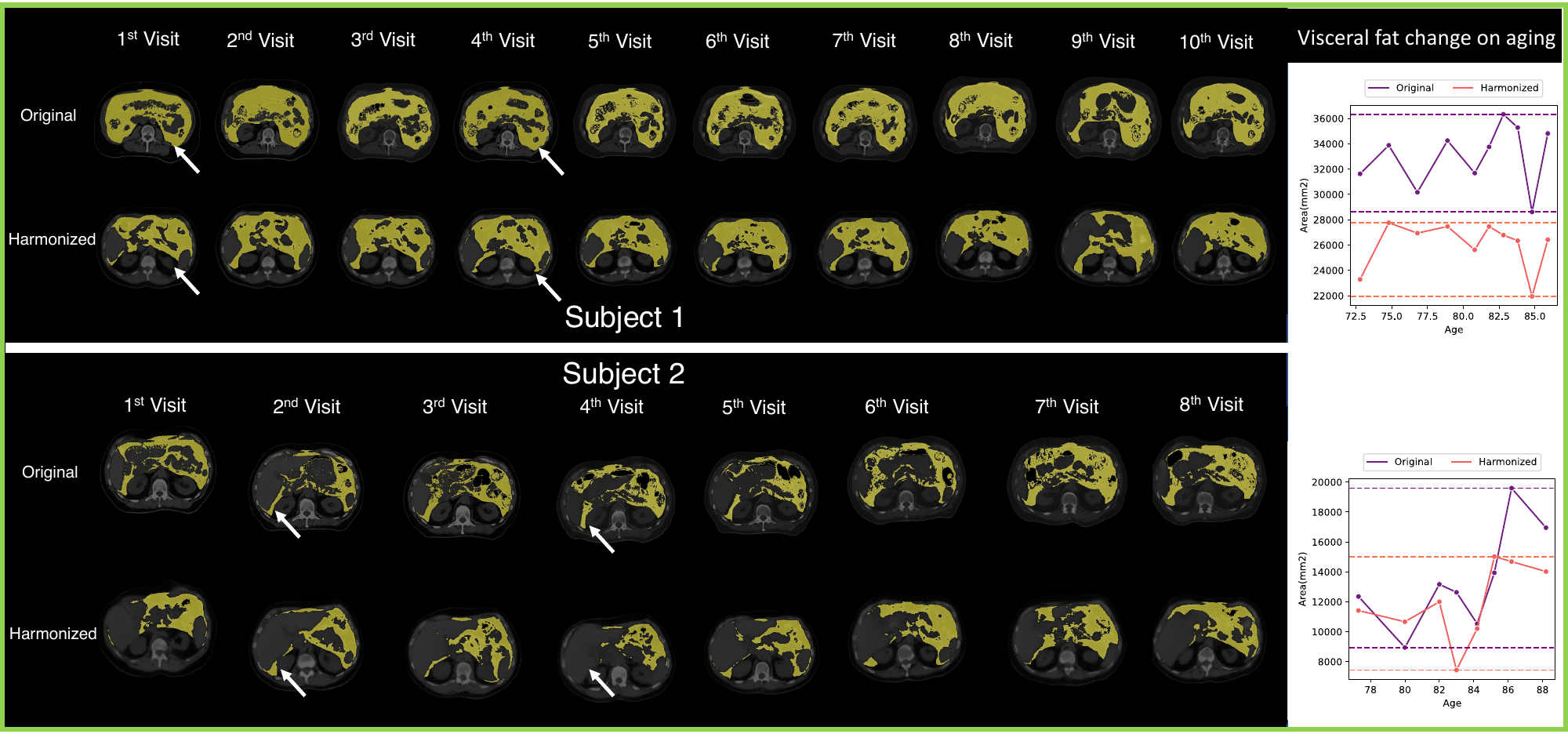}
\caption{Two subjects in the BLSA dataset that have 8 or 10 repeated visits. The yellow mask represents visceral fat. The rightmost column shows the visceral fat area change on aging. Similar to Fig.~\ref{fig:blsa_vis_short}, we highlight the image that is captured in different vertebral levels with white arrows. Subjects with a larger amount of repeated visits are more likely to have images captured at various vertebral levels. }
\label{fig:blsa_vis_long}
\end{figure*}

We calculate the NML on every two scans of the same subjects on both original and harmonized images, the results are shown in Fig.~\ref{fig:ml}(a). According to the result, the harmonized images have higher NML compared with the original images which indicates the generated images of the same subjects share a higher degree of similar information compared to those of the original images. We further evaluate with the CV, while we observe higher CV in the harmonized images compared with the original images, as it shown in Fig.~\ref{fig:ml}(b), which implies that differences between slices are increased after generation. As the metrics show contradictory results, we conduct human assessment of the harmonization results. We find that 431 out of 1033 subjects have at least one scan that is taken in obviously different vertebral levels compared with the other scans. For those 431 subjects, our model can help harmonize the positional variation and resulting a significantly lower CV than the original images with $p < 0.01$ under Wilcoxon signed-rank test, as it shown in Fig.~\ref{fig:cv}. Our models are able to effectively reduce the positional variance in subjects with both a lesser and greater number of longitudinal visits, as demonstrated in Fig.~\ref{fig:blsa_vis_short} and Fig.~\ref{fig:blsa_vis_long}, respectively. In Fig.~\ref{fig:blsa_vis_short}, our model reduce the variance by 36.3\% and 42.5\%, respectively. And in Fig.~\ref{fig:blsa_vis_long}, the variance is reduced by 37.8\% and 76.9\%, respectively. However, for those subjects that the original slice already taken at a similar vertebral level, our model can introduce additional noise and result in larger variance among scans, as it shown in Fig.~\ref{fig:blsa_limitation}. 

\subsection{Ablation Study}
\subsubsection{Adversarial regularization}
We compare our models trained with different $\beta$ scores to evaluate the impact of adversarial regularization. 
As it shown in Fig.~\ref{fig3}, when $\beta = 0.01$, the generated images are realistic and similar to the target slices. When $\beta=0$, there is no adversarial regularization, resulting in blurry generated images. By comparing the results with $\beta = 0$ and $\beta =0.01$ in Fig.~\ref{fig3}, it can be inferred that the adversarial regularization greatly improves the quality of the generated images.

\begin{figure*}[h!]
\centering
\includegraphics[width=\textwidth]{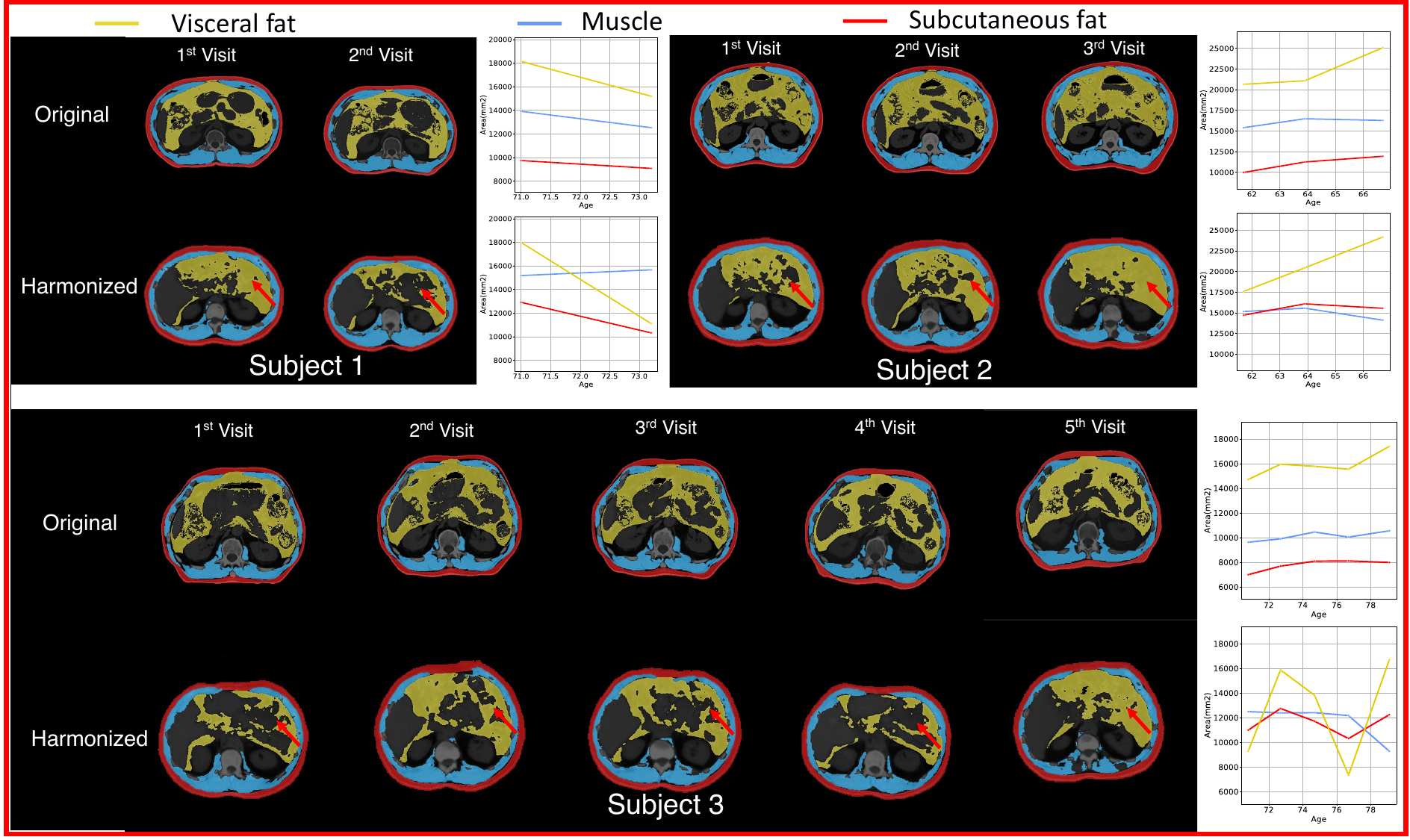}
\caption{Visualization of the original and harmonized images on the subjects that have scans taken at a similar vertebral level. Yellow, blue, and red masks represent visceral fat, muscle, and subcutaneous fat, respectively, which are generated by automatic segmentation methods. When the input images are at a similar vertebral level already, our model can introduce additional noise by predicting different heterogeneous soft tissues, as it highlighted with the red arrows. }
\label{fig:blsa_limitation}
\end{figure*}

This human qualitative assessment is aligned with LPIPS and NML results, as it shown in Table~\ref{tab:my-table}. However, it is different from the observation of the SSIM and the PSNR score, which are higher with $\beta=0$. This observation supports that SSIM and PSNR scores may not completely reflect human perception, as mentioned in~\cite{ledig2017photo,almalioglu2020endol2h}.

% This finding is consistent with the LPIPS results in Table.~\ref{tab:my-table}. However, the human qualitative assessment and LPIPS differ from the SSIM and PSNR as shown in Table.~\ref{tab:my-table} where SSIM and PSNR have higher scores with $\beta = 0$ on both dataset. This supports that SSIM and PSNR score may not fully represent a human perceptional assessment of image qualify, as it mentioned in \cite{ledig2017photo,almalioglu2020endol2h}. As for the longitudinal data harmonization, according to Fig.~\ref{fig4}, before applying our model, both muscle and visceral fat area have large fluctuations. These fluctuations have been reduced after mapping the slices to a similar vertebral level with our model C-SliceGen.

\begin{figure*}[h!]
\centering
\includegraphics[width=\textwidth]{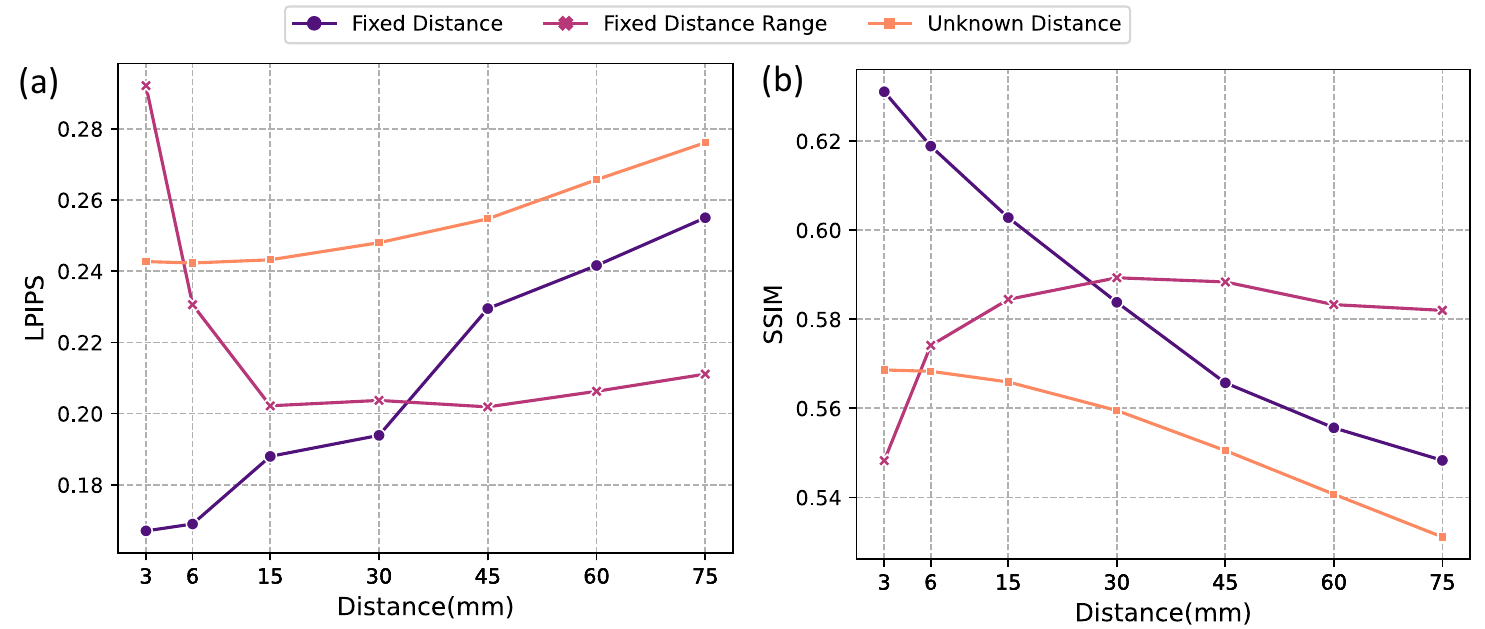}
\caption{Assessing the impact of the distance between the given slice and target slices on the model performance. Fixed distance represents the model trained with a known distance, and fixed distance range line represents the model trained with up to a distance of the given point. Unknown distance represents the inference performance of the model trained on the entire abdominal region when tested on data with a fixed distance between the input and target slices. We observe that the model's most effective distance range is between 15mm and 60mm where performance is stable in terms of both LPIPS and SSIM. }
\label{fig:spacing}
\end{figure*}

\subsubsection{Distance Impact}
In our method, we aim to map an abdominal axial scan at an arbitrary vertebral level to a pre-defined target vertebral level, where the distance between the given scan and the target scan is unknown. This is also the case in the BLSA single-slice dataset. We assume that as the distance difference between the scans increases, the scans will undergo more structural changes, making the generation process more challenging. To evaluate the impact of distance on our model performance, we conduct validation experiments. Specifically, we assess the performance of models trained on scans from varying distance ranges and compare them to models trained using known distances. We also include the model trained with the abdominal region with unknown distance for comparison (Model in Table~\ref{tab:my-table}). All the models are trained and tested with $\beta=0.01$, using the in-house portal venous dataset and BPR-based target slice selection method. We evaluate the model performance with the LPIPS and SSIM scores. The results are shown in Fig.~\ref{fig:spacing}.

To assess the performance of the model with fixed known spacing, we do not ask the model to predict the target slice since in most of the CT volumes, the spacing in the z dimension is 3mm. In this case, if we train with a fixed distance of 3mm, we can only get 2 conditional and target slice pairs for each subject, which lead to a data deficiency problem and cannot evaluate the model performance properly. Therefore, instead of predicting the target slice, we design to make each slice generate its corresponding slices that are 3mm, 6mm, and up to 75mm up in the abdominal region. For the model trained with a fixed spacing range, when the spacing is equal to 3mm, 6mm, and up to 75mm, which means the models are trained with the slices up to the specific distance to the target slice, respectively. The line unknown distance refers to the results by applying the trained model in Table~\ref{tab:my-table} to a fixed distance test set.  

According to Fig.~\ref{fig:spacing}, when the model has a fixed distance range of 3mm and 6mm, it has the worst performance among the other distance range. This can be explained by the data deficiency problem as mentioned before. The performance of models trained with fixed distance has the best performance when the slice distance is small but drops gradually with the distance increase, which indicates that the model does perform better with a smaller distance between the given and target slice. The models trained using the whole abdominal region (unknown spacing) consistently performed the worst, starting from a distance of 6mm. On the other hand, the models trained using a fixed range of 15mm to 60mm showed stable performance in terms of LPIPS and SSIM. These results indicate that the model's most effective distance range is between 15mm and 60mm, and training with a wider range can lead to decreased overall performance.

\section{Limitation and Discussion}
In this work, we improve the domain shift problem we observed in our previous publication~\cite{yu2022reducing} that when applying the model trained with the portal venous phase to the non-contrast phase BLSA data, the model has limited performance. We manage to reduce this issue by using non-contrast 3D volumetric data for training together with semi-BPR-based target slice selection for accurate target slice selection. From Fig.~\ref{fig:ml} and Fig.\ref{fig:cv}, we observe that our model can help reduce the longitudinal variance on data that are taken at obviously different vertebral levels. However, when it is applied to the data that are at similar vertebral levels, our model can introduce additional noise by predicting different heterogeneous soft tissues, as illustrated in Fig.~\ref{fig:blsa_limitation}. Predicting and synthesizing heterogeneous soft tissues such as the colon and stomach is challenging because these tissues' size and shape are largely dependent on individual conditions and position at the time of the CT scan, making it hard for the generative model to find a subject-specific distribution of such tissues. Hence, this remains a critical limitation of our current study. Solving the heterogeneous soft tissue, shape, and size generation problem can be the future work direction. Exploring solutions for the generation of heterogeneous soft tissue and preserving shape and boundary information can be the future work direction.

Additionally, we validate our method's most effective distance range. Models trained with data that are up to 60mm away from the target slice have comparable performance to that of the model trained with up to 15mm away from the target slice. Furthermore, the model trained within 60mm range performs much better than the model presented in Table~\ref{tab:my-table}, which is trained using slices from the entire abdominal region. Therefore, for the model to be most effective, it would be preferable to collect data within a range of no more than 60mm, or roughly around $\pm3$ vertebral level~\cite{kayalioglu2009vertebral} in future data collection.

% As the first work to use one abdominal slice to generate another slice, our approach currently has several limitations. (1) In most cases, the model is able to identify the position of each organ, but shape and boundary information are not well preserved. (2) It is hard to synthesize heterogeneous soft tissues such as the colon and stomach. (3) There is domain shift when the model trained on Portal Venous phase CT is applied to CT acquired in other phases 
% such as the non-contrast BLSA data.

\section{Conclusion}
% In this paper, we introduce our C-SliceGen model that conditions on an arbitrary 2D axial abdominal CT slice and generates a subject-specific slice at a target vertebral level. Our model is able to capture organ changes between different vertebral levels and generate realistic and structurally similar images. We further validate our model's performance on harmonizing the body composition measurements fluctuations introduced by positional variance on an external dataset. Our method provides a promising direction for handling imperfect single slice CT abdominal data for longitudinal analysis. \

Herein, we present our C-SliceGen model, which utilizes an arbitrary 2D axial abdominal CT slice as input and generates a subject-specific slice at a desired vertebral level. Our model can effectively capture changes in the organs across different vertebral levels and generate images that are realistic and structurally similar. Additionally, we demonstrate our model's effectiveness in harmonizing longitudinal body composition variance caused by positional differences among different visits in the BLSA single-slice CT dataset. This approach offers a promising solution for managing imperfect single-slice CT abdominal data in longitudinal analysis.

\section*{Declaration of generative AI in scientific writing 
}
During the preparation of this work the authors used chatGPT in order to improve readability and language. After using this tool, the authors reviewed and edited the content as needed and take full responsibility for the content of the publication. 

\section*{Disclosures}
The authors of the paper are directly employed by institutes or companies provided in this paper. No conflicts of interest exist in the submission of this paper.

\section*{Data Statement}
The in-house and BLSA data is unavailable to the public due to the sensitive nature of the research. The BTCV data is publicly online at: \href{https://www.synapse.org/#!Synapse:syn3193805/wiki/217789}{BTCV Website}.

% \href{https://github.com/Project-MONAI/model-zoo/tree/dev/models/wholeBrainSeg_Large_UNEST_segmentation}{MONAI Model}.
\section*{Acknowledgments}
This research is supported by NSF CAREER 1452485, 2040462 and the National Institutes of Health (NIH) under award numbers R01EB017230, R01EB006136, R01NS09529, T32EB001628, 5UL1TR002243-04, 1R01MH121620-01, and T32GM007347; by ViSE/VICTR VR3029; and by the National Center for Research Resources, Grant UL1RR024975-01, and is now at the National Center for Advancing Translational Sciences, Grant 2UL1TR000445-06. This research was conducted with the support from the Intramural Research Program of the National Institute on Aging of the NIH. The content is solely the responsibility of the authors and does not necessarily represent the official views of the NIH. The identified datasets used for the analysis described were obtained from the Research Derivative (RD), database of clinical and related data. The in-house imaging dataset(s) used for the analysis described were obtained from ImageVU, a research repository of medical imaging data and image-related metadata. ImageVU and RD are supported by the VICTR CTSA award (ULTR000445 from NCATS/NIH) and Vanderbilt University Medical Center institutional funding. ImageVU pilot work was also funded by PCORI (contract CDRN-1306-04869).

\bibliography{report}   % bibliography data in report.bib
\bibliographystyle{spiejour}   % makes bibtex use spiejour.bst

%%%%% Biographies of authors %%%%%

% \vspace{2ex}\noindent\textbf{First Author} is an assistant professor at the University of Optical Engineering. He received his BS and MS degrees in physics from the University of Optics in 1985 and 1987, respectively, and his PhD degree in optics from the Institute of Technology in 1991.  He is the author of more than 50 journal papers and has written three book chapters. His current research interests include optical interconnects, holography, and optoelectronic systems. He is a member of SPIE.

% \vspace{1ex}
% \noindent Biographies and photographs of the other authors are not available.

\listoffigures
\listoftables

\end{spacing}
\end{document}